# Exploring the distribution of connectivity weights in resting-state EEG networks

Shiang Hu[1*], Xiao Gong[1], Xiaolong Huang[1], Jie Ruan[1*], Pedro Antonio Valdes-Sosa[2, 3]

[1]Anhui Province Key Laboratory of Multimodal Cognitive Computation, School of Computer Science and Technology, Anhui University, 111 Jiulong Road, Hefei, 230601, China

[2]The Clinical Hospital of Chengdu Brain Science Institute, MOE Key Lab for Neuroinformation, School of Life Science and Technology, University of Electronic Science and Technology of China, Chengdu, China

[3]Cuban Center for Neurocience, La Habana, Cuba

Corresponding author:　　shu@ahu.edu.cn (Shiang Hu), JieRuan97@outlook.com (Jie Ruan)

## Abstract

The resting-state brain networks (RSNs) reflects the functional connectivity patterns between brain modules, providing essential foundations for decoding intrinsic neural information within the brain. It serves as one of the primary tools for describing the spatial dynamics of the brain using various neuroimaging techniques, such as electroencephalography (EEG) and magnetoencephalography (MEG). However, the distribution rules or potential modes of functional connectivity weights in the resting state remain unclear. In this context, we first start from simulation, using forward solving model to generate scalp EEG with four channel densities (19, 32, 64, 128). Subsequently, we construct scalp brain networks using five coupling measures, aiming to explore whether different channel density or coupling measures affect the distribution pattern of functional connectivity weights. Next, we quantify the distribution pattern by calculating the skewness, kurtosis, and Shannon entropy of the functional connectivity network weights. Finally, the results of the simulation were validated in a normative database. We observed that: 1) The functional connection weights exhibit a right-skewed distribution, and are not influenced by channel density or coupling measures; 2) The functional connection weights exhibit a relatively uniform distribution, with the potential for volume conduction to affect the degree of uniformity in the distribution; 3) Networks constructed using coupling measures influenced by volume conduction exhibit significant correlations between the average connection weight and measures of skewness, kurtosis, and Shannon entropy. This study contributes to a deeper understanding of RSNs, providing valuable insights for research in the field of neuroscience, and holds promise for being associated with brain cognition and disease diagnosis.

# 1. Introduction

The network is a collection of nodes and links between nodes, serving as a mathematical representation of relational information among populations[1], characterized by vast and intricate multidimensional spatial systems. Brain networks typically utilize electrodes (on the scalp) or brain regions (on the cortex) as vertices, with the coupling strength between signals serving as link weights, interpreted as the influx or efflux of information flow between nodes. The brain network reveals the spatial dynamics of the brain, which manifest as connectivity patterns among functional modules of the brain[2]. It finds extensive applications in human cognition and behavior, as well as in the diagnosis of psychiatric disorders, such as the establishment of normative brain network evolution across the lifespan[3], emotion recognition[4], and the exploration of biomarkers for Alzheimer's disease[5] and depression[6], etc.

In the process of analyzing brain networks, researchers commonly employ thresholding techniques to simplify the intricate connectivity patterns due to their complex nature. The overarching principle of thresholding is to disregard weak connections while preserving strong ones. Thresholding methods include absolute thresholding, proportional thresholding[7], and double thresholding[8]. In absolute thresholding, the threshold value remains fixed and does not change with variations in conditions. Proportional thresholding involves setting a fixed proportion value $p \in (0, 1)$, and subsequently preserving a proportion p of the total connections, making the thresholding network density consistent across different conditions, hence it is also referred to as density-based thresholding. Double thresholding initially employs absolute thresholding to obtain a network T1 with strong connections and discards weak connections to form T2. Subsequently, the average of non-empty connections in each row of T2, along with their standard deviation, is calculated to determine the second threshold. This second threshold is then applied to T2 to derive a second thresholding network, T3. Finally, the network is represented as the union of T1 and T3.

Despite the significant advancements achieved in thresholding-based analysis of brain networks, the loss of weak connections undoubtedly represents an information deficit in brain network analysis. These weak connections may play crucial roles in the overall network functionality, and their loss could lead to an incomplete understanding of brain network functions, even in the context of resting-state networks. We consider that the distribution of connectivity weights in spontaneous resting-state EEG brain networks follows a certain statistical regularity, a question that has remained unanswered hitherto. The purpose of this paper is to investigate the distribution rules of connection weights in fully connected resting-state networks, aiming to uncover their underlying distributional patterns.

There are numerous factors that influence the functional connectivity of brain networks. Specifically, noise can weaken functional connections within the network, reducing its overall connectivity[9, 10]. The loss of brain components generates a low-rank signal matrix, while coupling between channels is enhanced, leading to strengthened network connections[11, 12]. Additionally, the number of channels, representing nodes in the network, undoubtedly impacts the functional connectivity of the network[13, 14]. An easily overlooked significant influencing factor is the measurement of functional connectivity in brain networks. Research indicates that different measurement methods exhibit variances among the investigated network connections, such as the effects of volume conduction, sensitivity to noise and loss of brain activity[11, 15].

The quantification of the distribution of brain network functional connectivity weights typically involves statistical measures such as the mean, median, mode, and standard deviation[16]. More complex statistical measures like skewness and kurtosis are used to quantify the shape of the data distribution. Additionally, some research has proposed the use of Shannon entropy (SE) to quantify the uniformity of data distribution. This method is employed to estimate whether the distribution of network connectivity weights is clustered or uniform[17].

In this study, we explore the distribution patterns of brain network functional connectivity weights starting from simulated and normative data. In the simulation, to investigate whether the distribution pattern is constrained by electrode density and functional connectivity metrics, we respectively generated simulated EEG data with 19, 32, 64, and 128 channels, and constructed brain networks using coupling metrics such as coherence, the imaginary part of coherency, phase-locking value, phase-lag index, and amplitude envelope correlation. Subsequently, statistical analyses of the weight distributions of each combination (channel × coupling metric) of brain networks were conducted, with computed metrics including mean, skewness, kurtosis, and SE. Pearson correlations were then calculated for pairs of mean-skewness, mean-kurtosis, and mean-SE. The data generation and analysis workflow are depicted in **Fig. 1**.

## 2. Materials and methods

### 2.1 Simulation

#### 2.1.1 Intracranial EEG

To ensure that the simulated data exhibit realistic spectral characteristics of EEG, cortical activity was modeled using real intracranial EEG (iEEG) signals[18]. The Montreal Neurological Institute and Hospital (MNI) has released iEEG data extracted from normal brain regions of 106 epilepsy patients, accumulating a total of 1785 channels. After removing bad channels, 1772 remaining channels of iEEG data cover the entire brain, providing an atlas of human EEG patterns[19].

The cortical surface consisted of 15,002 dipoles aligned with all cortical vertices. For computational convenience, this surface was downsampled to 3002 vertices. Subsequently, 200 signal sources were randomly selected from the iEEG data containing 1772 channels, with each source activity comprising 10,000 samples. The remaining 2802 signal sources were simulated as low-intensity Gaussian noise to ensure activity across all dipoles on the cortical surface. Finally, these 3002 signal sources were shuffled to increase randomness, resulting in the generation of dipole activities with dimensions of 3002 × 10000.

### 2.1.2 Forward solution

To obtain scalp EEG, solving the forward EEG problem is necessary. The head model was generated from the ICBM 152 MRI template and used for computing the gain (lead field) matrix. The gain matrix was obtained using the boundary element method, a technique embedded in the OpenMEEG[20] package integrated into Brainstorm[21]. The scalp electrodes were setup from the 10-20 system with 19 channels and the GSN HydroCel (EGI, electric geodesic Inc) electrode configuration systems with 32, 64, and 128 channels. As a result, the dimensions of the resulting lead field matrix were 19/32/64/128 × 3002. The estimated scalp EEG representation through forward solution is as follows:

$$V(t) = (v_1(t) \cdots v_{N_c}(t))^T = G \cdot (x_1(t) \cdots x_{N_s}(t))^T = GX(t) \quad (1)$$

where $V_t$ represents scalp EEG, $N_c$ denotes the number of electrodes, $t$ stands for the time sample points, $G$ signifies the lead field matrix, $N_s$ represents the number of dipoles, and $X(t)$ indicates cortical activity.

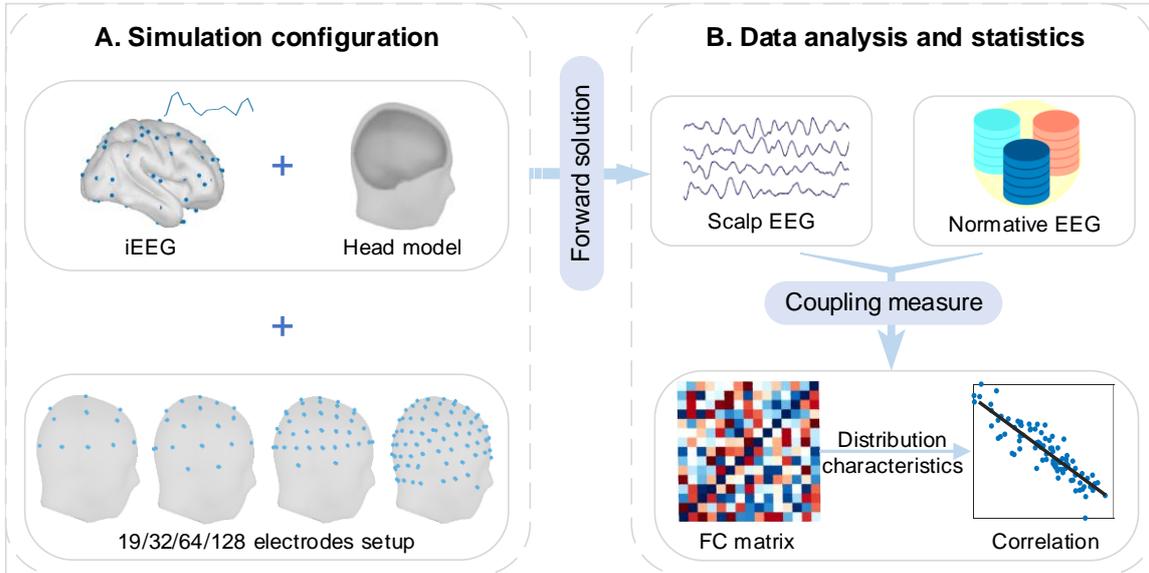

*Fig. 1. Schematic diagram of data generation and analysis.* A. Simulation configuration, including source model, head model, and electrodes setup; B. Data analysis and statistics, encompassing both simulated data and normative data. iEEG: intracranial EEG, FC: functional connectivity.

## 2.2 The normative EEG database

The normative EEG database is derived from a multicenter qEEG normative project[22], with data collected from 1966 participants across 9 countries and 14 sites. The data recordings utilized 19 channels based on the 10/20 International Electrode Placement System: Fp1, Fp2, F3, F4, C3, C4, P3, P4, O1, O2, F7, F8, T3/T7, T4/T8, T5/P7, T6/P8, Fz, Cz, and Pz. Each site conducted preprocessing on the EEG, excluding Independent Component Analysis (ICA). The cross-power spectra of scalp EEG were calculated using Bartlett's method by averaging the periodograms of more than 20 consecutive non-overlapping segments. The frequency range covered is from 1.17 to 19.14Hz with a resolution of 0.39Hz. The shared dataset includes the power spectral and cross-spectral data of EEG, as well as anonymized participant information, such as age and gender, along with technical parameters like EEG recording equipment, electrode placements, EEG recording references, laboratory details, and the country of data collection.

Due to one site among the 14 employing the Cz electrode as a reference, resulting in zero-cross spectrum data for this electrode. Additionally, connectivity patterns referenced to the Cz electrode exhibit significant distortion[23]. Average reference can alleviate connection distortion. Nonetheless, as a member of the single-level reference family, average reference introduces additional linear dependencies[22, 24], consequently enhancing network connectivity. Therefore, in this study, we discarded the data from this site, leaving a remaining 1791 cases. The distribution of these data, originating from nine countries, is illustrated in **Fig. 2**.

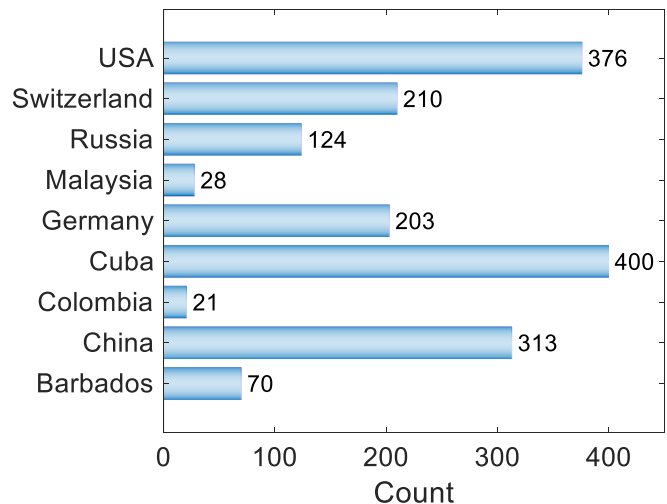

*Fig. 2. Overview of data distribution from nine countries.* *Note that this distribution represents the 1971 cases after excluding data from one site.*

## 2.3 Functional connectivity

After estimating scalp EEG with varying numbers of electrodes, scalp brain networks were constructed based on coupling metrics. Here, five coupling metrics were utilized, namely coherence, the imaginary part of coherency, phase-locking value, phase-lag index, and amplitude envelope correlation.

### 2.3.1 Coherence and the imaginary part of coherency

Coherency measures the linear relationship between signals at a specific frequency[25], while coherence measures the covariance of signals in the frequency domain. Assuming the time series of channels $i$ and $j$, denoted as $x_i(t)$ and $x_j(t)$ respectively, undergo Fourier transformation resulting in $x_i(f)$ and $x_j(f)$, the cross-spectrum is defined as:

$$S_{ij}(f) = <x_i(f)x_j^*(f)> \quad (2)$$

where $<\cdot>$ represents expectation value, $*$ means complex conjugation and $f$ is specific frequency. Then coherency is defined as the normalization of the cross-spectrum:

$$C_{ij}(f) = \frac{S_{ij}(f)}{\sqrt{(S_{ii}(f)S_{jj}(f))}} \quad (3)$$

magnitude square coherence (MSC) or coherence (COH) is the square of the coherency, and the imaginary part of coherency (iCOH) represents the imaginary information of coherency:

$$\begin{cases} COH_{ij}(f) = |C_{ij}(f)|^2 \\ iCOH_{ij}(f) = \Im(C_{ij}(f)) \end{cases} \quad (4)$$

where $|\cdot|$ represents the absolute value and $\Im$ denotes the imaginary part of $C_{ij}(f)$.

### 2.3.2 Phase-locking value

Phase-locking value (PLV), a classical method for quantifying phase synchronization between signals, is achieved by calculating the phase difference between signals[26, 27]. The definition of PLV is as follows:

$$PLV = |<\exp(i(\Phi_1 - \Phi_2))>| \quad (5)$$

where $<\cdot>$ represents expectation value, $|\cdot|$ represents the absolute value, $i$ represents the imaginary unit, $\Phi_1$ and $\Phi_2$ denote the instantaneous phases of the signals.

### 2.3.3 Phase-lag index

Phase Lag Index (PLI) aims to address the volume conduction and reference electrode problems. It measures the asymmetry in the distribution of phase differences between two signals and is a pure phase measure[27, 28]. The definition of PLI is as follows:

$$PLI = |<sign(\Phi_1 - \Phi_2)>| \quad (6)$$

where $<\cdot>$ represents expectation value, $|\cdot|$ represents the absolute value, $\Phi_1$ and $\Phi_2$ denote the instantaneous phases of the signals.

### 2.3.4 Amplitude envelope correlation

Amplitude envelope correlation is a measure within the family of cross-frequency coupling approaches. It evaluates the amplitude interactions between signals by computing the Pearson correlation of their amplitude envelopes[29].

In this study, due to the absence of raw EEG in the normative data, only cross-spectral information was included. Therefore,

for the normative data, functional connectivity measures were limited to COH and iCOH, and comparisons were made across different frequency bands. However, for the simulated data, all five functional connectivity measures were utilized. Given that alpha band (8 - 13 Hz) connectivity exhibited the most prominent patterns[30], comparisons were restricted to the alpha band during the simulation phase. COH and iCOH were computed based on the normalized cross-spectrum. PLV, PLI, and AEC were computed over continuous sliding windows with a size of 6 seconds. For PLV, the window had no overlap, whereas for PLI and AEC, the windows overlap by 0.5 seconds. Finally, the average across all windows was taken to achieve the computation.

## 2.4 Connectivity weights distribution characteristics

All functional connectivity matrices constructed through the aforementioned coupling measurements are symmetric matrices. We extracted the upper triangle of the functional connectivity matrix and concatenate it into a one-dimensional vector and quantified the distribution using mean connectivity weights (MCW), skewness, and kurtosis, as well as the uniformity of the data. Detailed descriptions of each measure are provided below.

### 2.4.1 Skewness and kurtosis

Skewness is a measure of the degree of asymmetry in the statistical distribution of data, providing an indication of how the probability density curve deviates from symmetry relative to the mean. A positive skewness signifies that the shape of the data distribution has a longer tail on the right side of the mean, with the majority of the data concentrated on the left side of the mean, which is called right-skewed distribution. Conversely, a negative skewness indicates a longer tail on the left side, with the data concentrated on the right side of the mean, which is called left-skewed distribution. The skewness of a Gaussian distribution is zero. Kurtosis is another statistical measure used to describe the shape of a data distribution, quantifying the relative sizes of the tails and peak of the distribution. Specifically, kurtosis measures the sharpness of the peak in the data distribution. The kurtosis of a normal distribution is 3. Distributions with higher kurtosis values indicate that the data are more concentrated around the mean and have sharper tails. Conversely, distributions with lower kurtosis values suggest that the data are more spread out with flatter tails. The skewness and kurtosis of a data distribution is defined as follows,

$$\begin{cases} s = \dfrac{E(x-\mu)^3}{\sigma^3} \\ k = \dfrac{E(x-\mu)^4}{\sigma^4} \end{cases} \quad (7)$$

where $\mu$ is the mean of $x$, $\sigma$ is the standard deviation of $x$, and $E$ represents the expected value, $s$ and $k$ represent the skewness and kurtosis, respectively.

### 2.4.2 Shannon entropy

Shannon entropy (SE) is utilized to characterize the uniformity of the distribution of network connection weights[17], with a range between 0 and 1. A value closer to 0 indicates a more concentrated distribution, while a value closer to 1 signifies a more uniform distribution. The computation of SE is related to the relative probability histogram of network connection weights. For a given histogram with a specified number of bins, SE is defined as follows:

$$\text{SE} = \frac{-1}{\log_2 N} \sum_{i=1}^{N} P_i \log_2(P_i) \tag{8}$$

where $P_i$ represents the relative probability of each bin, $N$ indicates the number of histogram bins, and $\log_2 N$ is the normalization factor. Here, the range of connection weights is [0, 1]. To achieve finer segmentation, we set the number of bins to 100, with each bin having a width of 0.01.

## 2.5 Statistical analysis

To explore the distribution patterns of network connections, we conducted a statistical analysis of MCW, skewness, kurtosis, and SE of the networks. Pearson correlations were then analyzed pairwise between MCW-skewness, MCW-kurtosis, and MCW-SE, and the Pearson correlation coefficients along with their significance were reported, including both simulated and normative data.

## 3. Results

## 3.1 Simulation analysis

### 3.1.1 The network connection weights exhibit a right-skewed distribution

The combined effect of skewness and mean can quantify the symmetry of the distribution of network connection weights[12]. **Fig. 3** illustrated the findings of skewness and mean connectivity weights (MCW) for all combinations of electrode montage and connectivity metrics. It can be observed that the skewness in each subplot is greater than 0, which holds true for all combinations of electrode montage and connectivity metrics. This indicates a right-skewed distribution of network connectivity weights, unaffected by the number of channels and coupling measures. For each coupling measure, the MCW typically lies within the left subset of [0, 1], indicating that network connection weights are predominantly distributed towards lower values. For each electrode montage, the MCW increased with the number of channels, but the variation was marginal.

Additionally, in networks constructed using COH, PLV, and AEC, there exists a significant negative correlation (p < 0.001) between the skewness of connection weights and their mean values. Moreover, generally, the Pearson correlation coefficient tends to increase with the number of channels. It is noteworthy that such a correlation is not observed in networks constructed using iCOH and PLI.

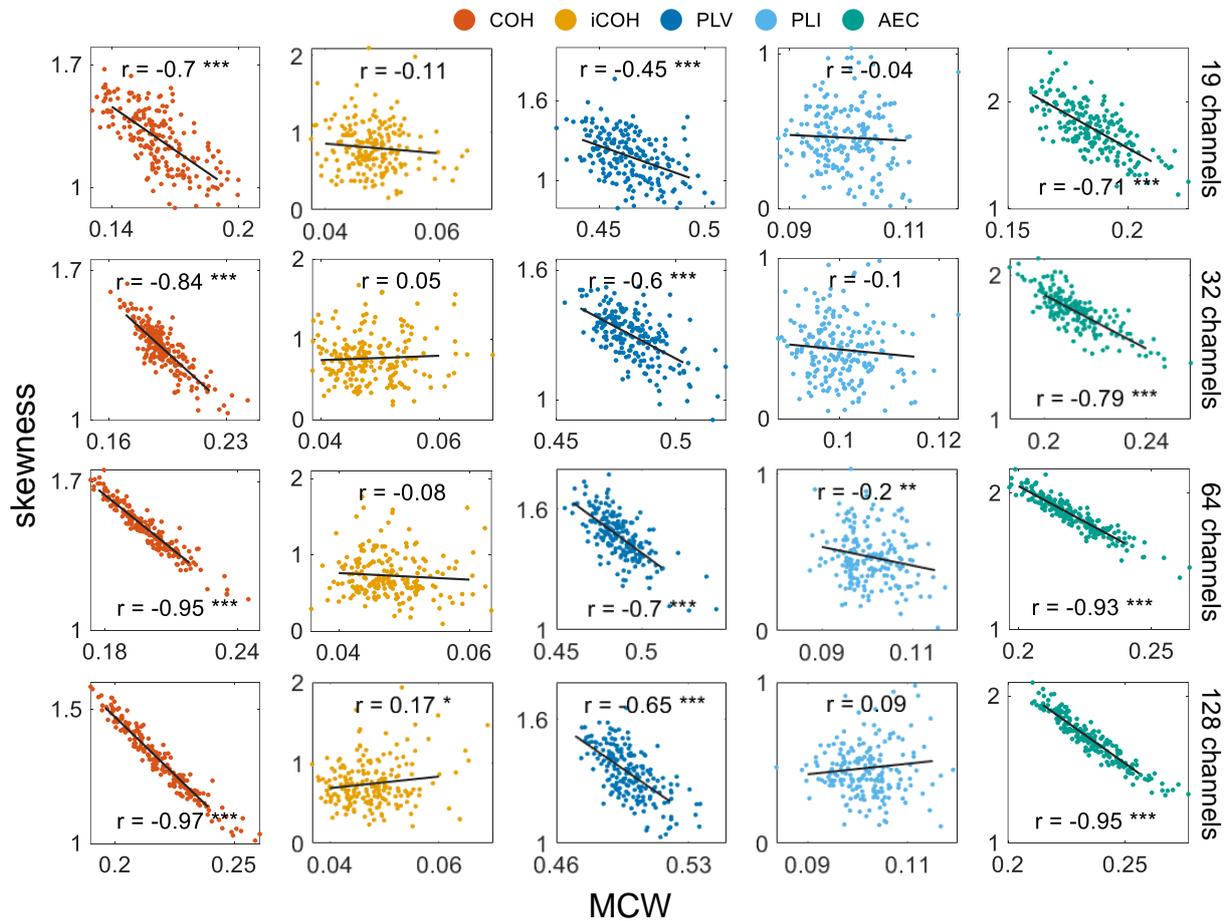

*Fig. 3. Scatter plots and correlations between skewness and mean connectivity weights (MCW) for all combinations of electrode montage and connectivity metrics. Each subplot is annotated with Pearson correlation coefficient and significance test level, where ** indicates p < 0.01, and *** indicates p < 0.001. Note that the absence of * indicates p > 0.05.*

### 3.1.2 The network connection weights display a leptokurtic distribution

Visually, kurtosis reflects the sharpness of the peak of a distribution, serving as a measure of the "tails" of the data, depicting the steepness of the distribution. The kurtosis of a normal distribution is 3. A kurtosis exceeding three indicates a sharper peak, suggesting a steeper shape compared to the peak of a normal distribution. Conversely, the opposite holds true. **Fig. 4** depicted the results of kurtosis under various combinations of electrode montage and connectivity metrics.

From **Fig. 4**, it can be observed that regardless of the electrode montage and connectivity metric combinations, the kurtosis is consistently less than 3. Moreover, for the coupling metrics iCOH and PLI, the kurtosis is even smaller. This indicates that compared to a normal distribution, the distribution of network connection weights is flatter, resembling a platykurtic distribution. Furthermore, for COH, PLV, and AEC, there exists a significant negative correlation between kurtosis and mean connectivity weights (p < 0.001), with a substantial correlation coefficient, which increases with the number of channels. However, there is no correlation between kurtosis and mean connectivity weights for iCOH and PLV.

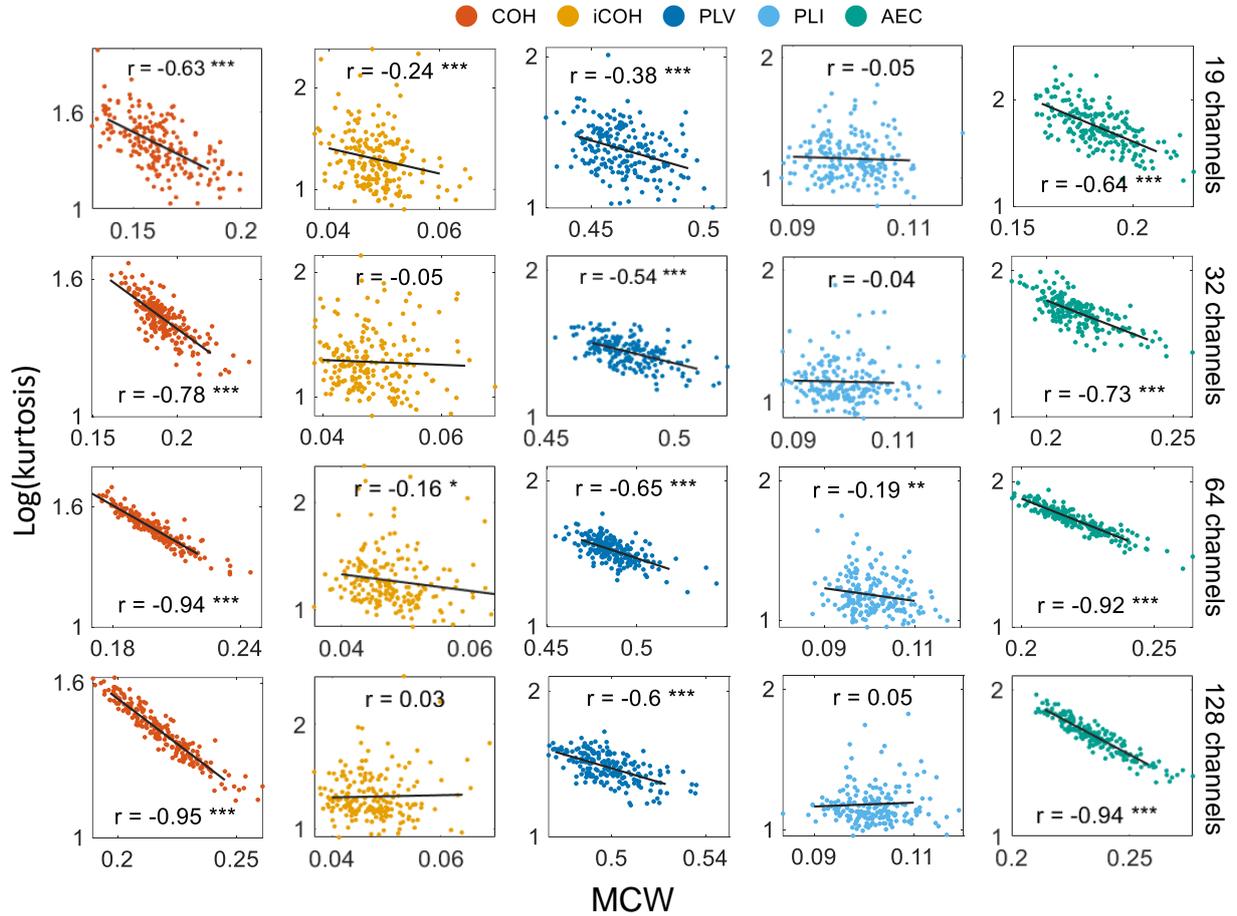

*Fig. 4. Scatter plots and correlations between kurtosis and mean connectivity weights (MCW) for all combinations of electrode montage and connectivity metrics. Each subplot is annotated with Pearson correlation coefficient and significance test level, where * indicates p < 0.05, and *** indicates p < 0.001. Note that the absence of * indicates p > 0.05.*

### 3.1.3 The network connection weights exhibit a relatively uniform distribution

Shannon entropy (SE) quantifies the uniformity of network connection weight distribution[17], with values ranging from 0 to 1. The greater the uniformity of the weight distribution, the higher the SE, and vice versa. **Fig. 5** illustrates the results of SE and MCW under various combinations of electrode montage and connectivity metrics.

Firstly, regardless of the combination of electrode montage and connectivity metrics, there was a significant positive correlation between MCW and SE (p < 0.001), indicating that larger MCW is associated with larger SE, and this correlation is independent of electrode count and coupling measures. Secondly, for coupling metrics COH, iCOH, and AEC, the Pearson correlation coefficients are around 0.9 and increase with the number of channels. However, for PLV and PLI, the Pearson correlation coefficients are relatively lower. The commonality between PLV and PLI lies in their reliance on signal phase information. Next, For COH, PLV, and AEC, the values of SE fall within the range of 0.7 to 0.9, indicating that networks constructed with these coupling measures exhibit a uniform distribution of connection weights. For iCOH and PLI, SE values range from 0.3 to 0.55, suggesting a relatively lower degree of uniformity in the distribution of connection weights.

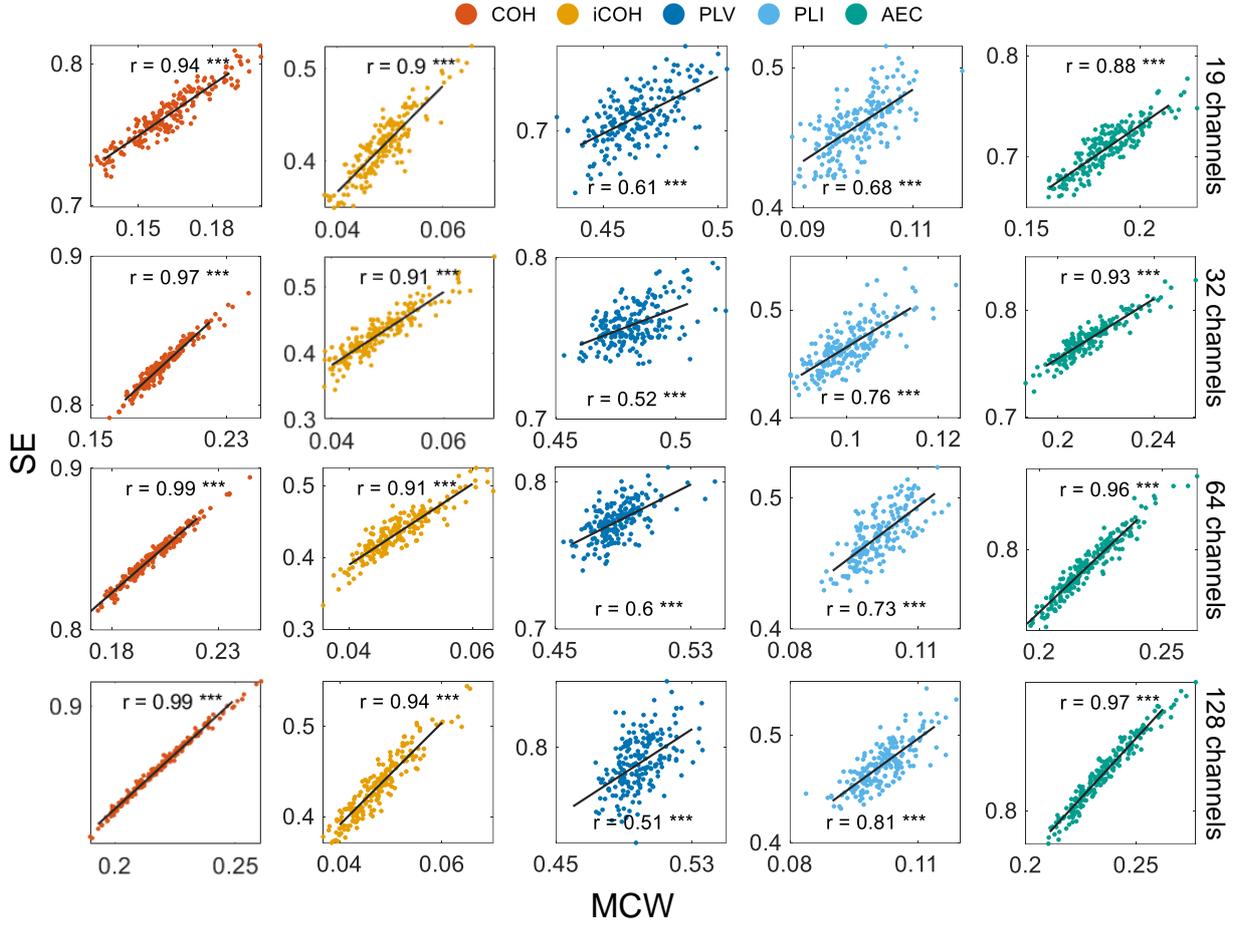

*Fig. 5. Scatter plots and correlations between Shannon entropy (SE) and mean connectivity weights (MCW) for all combinations of electrode montage and connectivity metrics. Each subplot is annotated with Pearson correlation coefficient and significance test level, where \*\*\* indicates p < 0.001.*

## 3.2 Validation in normative EEG

Due to the lack of pure EEG as the ground truth[31], here, we validated using publicly available normative datasets[22]. These datasets include cross-spectra, which serve as the basis for computing COH and iCOH. While the mathematical relationship between cross-spectra and coupling measures such as PLV and PLI has been derived, this relationship is confined to electrophysiological recordings with Gaussian distributions[27]. Thus, validation in normative data is limited to COH and iCOH.

The validation results of COH and iCOH in the normative data are depicted in **Fig. 6** and **Fig. 7**. Each column from left to right represents the results for four frequency bands, with brightness in each subplot indicating the degree of data clustering. From the results, it can be observed that there is no significant difference among the four frequency bands, and skewness/kurtosis/SE are significantly correlated with MCW.

From the first row of **Fig. 6**, it can be observed that the skewness is predominantly greater than 0, with only a few instances below 0. This suggests that the weights of the network constructed by COH exhibit a right-skewed distribution, which aligns with the simulation results. It is noteworthy that, compared to delta and theta bands, the alpha band exhibits smaller skewness and higher MCWs, whereas the beta band shows greater skewness and lower MCWs. This suggests that, in the alpha band, the overall network connectivity weights are biased towards larger values, which may be associated with heightened alpha rhythm activity during resting-state EEG. The second row displays the results of kurtosis in weight distribution. Consistent with the simulation results, there exists a significant negative correlation between kurtosis and mean connection weight. Similarly, compared to the delta and theta bands, the kurtosis of the alpha band is smaller, while the kurtosis of the beta band is larger. In comparison to the normal distribution, the network weight distribution in the alpha band tends toward a flat peak, while in the beta band, it tends toward a sharper peak. For the delta and theta bands, the network weight distribution is moderate. SE reflects the uniformity of network connection weights distributed in the interval [0, 1]. The third row of **Fig. 6** shows the results of SE. Compared to delta and theta bands, the SE of the alpha band is larger, while that of the beta band is smaller. This phenomenon is opposite to the results of skewness and kurtosis. This implies that the network connection weights in the alpha band are more evenly distributed, those in the delta and theta bands are relatively moderate, and those in the beta band are the least uniform. Overall, the SE values for each band are generally greater than 0.7, indicating that the network connection weights exhibit a relatively uniform distribution.

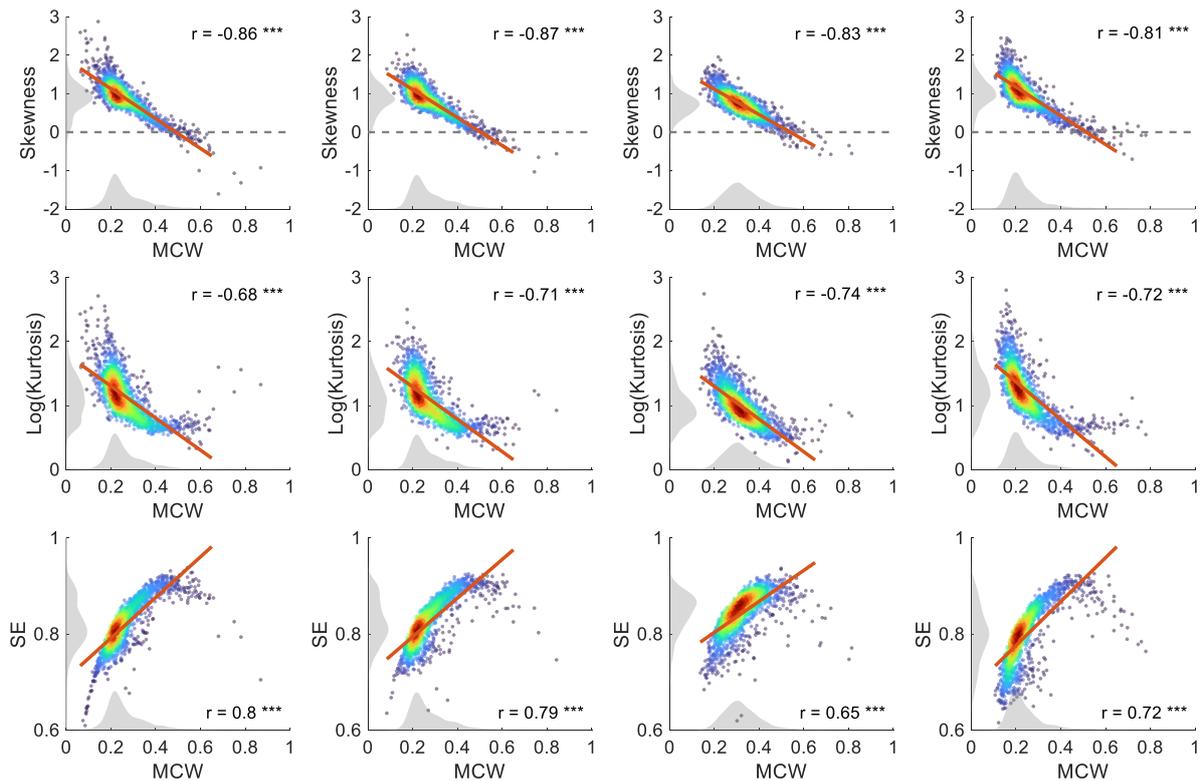

*Fig. 6. Validation of COH in normative data. Each row, from top to bottom, presents the results of skewness, kurtosis, and Shannon entropy (SE), while each column, from left to right, represents four frequency bands: delta, theta, alpha, and beta. The brightness of each scatter plot in each subplot indicates the degree of data clustering, while the gray shaded area on the axes represents the distribution of corresponding values. r denotes the Pearson correlation coefficient, and \*\*\* indicates p < 0.001.*

**Fig. 7** illustrates the results of the network connectivity weight distribution constructed by iCOH. Compared to networks constructed by COH, those built by iCOH exhibit a generally similar network connectivity weight distribution. Unlike what is observed in **Fig. 6**, the skewness and kurtosis of the network connectivity weight distribution constructed by iCOH are more concentrated (grey shaded area). The skewness and kurtosis exhibit a smaller Pearson correlation coefficient with MCW, similar to the simulation results. Additionally, an intriguing phenomenon is observed in the delta, theta, and beta bands, where the distribution of MCWs exhibits two peaks, indicating clustering at two distinct positions. The SE of the network connectivity weight distribution constructed from COH and iCOH exhibits differences, specifically with iCOH demonstrating relatively lower value. This implies a more nonuniform distribution of connection weights. However, the Pearson correlation coefficient between SE and MCW is larger.

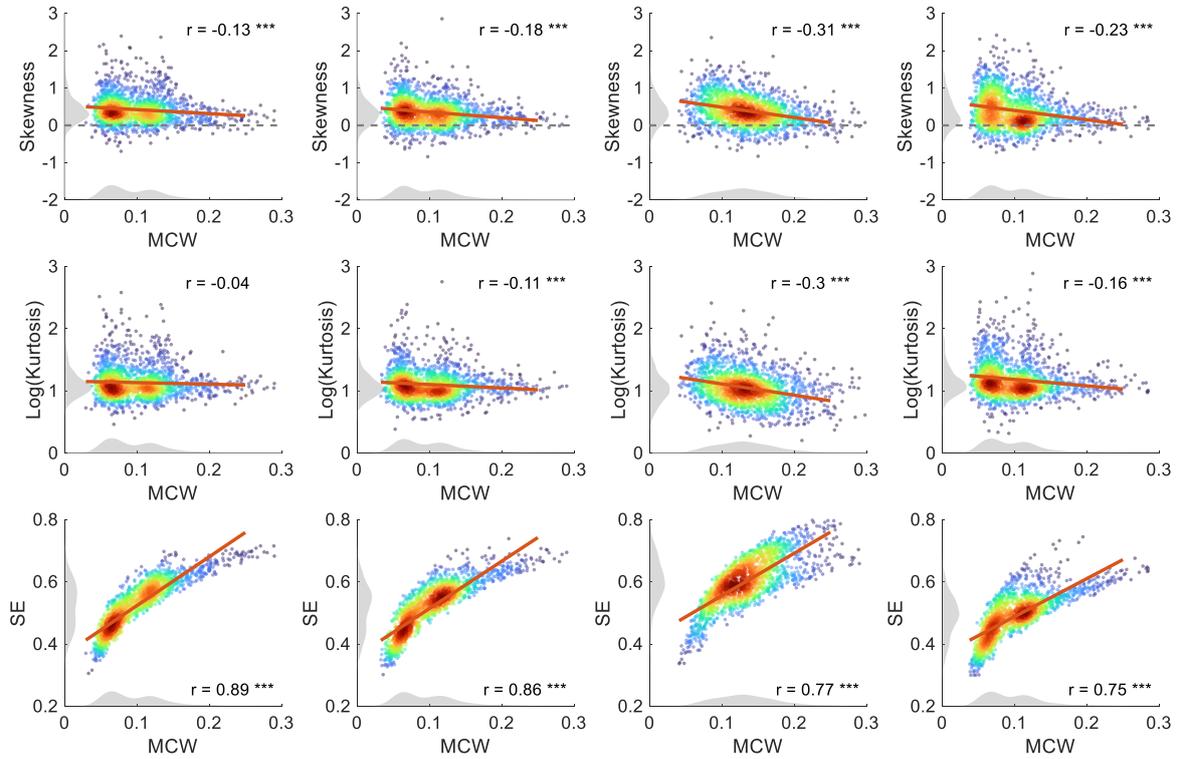

*Fig. 7. Validation of iCOH in normative data.* Each row, from top to bottom, presents the results of skewness, kurtosis, and Shannon entropy (SE), while each column, from left to right, represents four frequency bands: delta, theta, alpha, and beta. The brightness of each scatter plot in each subplot indicates the degree of data clustering, while the gray shaded area on the axes represents the distribution of corresponding values. r denotes the Pearson correlation coefficient, and *** indicates $p < 0.001$. Note that the absence of * indicates $p > 0.05$.

## 4. Discussion

The present study comprehensively investigates the distribution patterns of functional connectivity weights in the resting-state brain networks, both through simulation and normative data validation. This includes exploring potential factors influencing these distribution patterns, such as the number of electrodes (19, 32, 64, 128) and coupling measures (COH, iCOH, PLI, PLV, AEC). Our findings indicate that the functional connectivity weights in the resting-state brain networks exhibit a right-skewed distribution, which is independent of channel density and the coupling measure used. Networks constructed using coupling measures unaffected by volume conduction exhibit a more uniform distribution of network connection weights compared to those affected by volume conduction. These results provide deep insights into resting-state brain networks and may inform various related endeavors, including quality control, emotion recognition, and psychiatric disorder identification.

Skewness and kurtosis have been incorporated in numerous investigations, such as seizure detection[32, 33] and localization in epilepsy[34], assessment of functional network robustness in Alzheimer's disease[35], emotion recognition and diagnosis of post-traumatic stress disorder [36], classification of EEG signals in neural networks[37], as well as artifact removal and EEG signal quality control[38–40]. Additionally, in a study exploring the network distribution in Alzheimer's disease patients, Shannon entropy (SE) has been proven to be an effective method for characterizing the distribution of network weights[17]. In this study, kurtosis, skewness, and SE are utilized as quantitative features of network connectivity weight distributions. Skewness and kurtosis, as the third and fourth standardized moments of sample data, quantify the shape of the probability density function, and SE quantifies the uniformity of the distribution of random variables within a known interval.

As is well known, a major factor influencing the functional connectivity of brain networks is channel density. Nodes, serving as the physical representation of electrodes in the network, may directly influence the distribution pattern of functional connection weights. To date, several studies have investigated the impact of channel density on the analysis of functional brain network connectivity. More specifically, source-level functional connectivity analysis should ensure a minimum of 64 scalp EEG electrodes to infer cortical dynamics from scalp EEG signals[13, 14]. This study employs forward solution to simulate scalp EEG signals with channel densities of 19, 32, 64, and 128, and constructs scalp brain networks, aiming to investigate whether channel density influences the distribution pattern of network weights. From **Fig. 3**, **Fig. 4** and **Fig. 5**, it appears that increasing the number of electrodes may lead to larger Pearson correlation coefficients and MCW values, yet it does not affect the overall distribution of connection weights. This can be easily comprehended, as the augmentation of network nodes results in a greater number of connection weights, yet these weights may still adhere to the original distribution rule.

Another key factor that may influence the distribution of network connection weights is the coupling metric used in constructing the network. Thus far, a plethora of network connectivity measures have been proposed, all developed with the aim of characterizing interactions within brain networks to uncover cognitive and behavioral information[41, 42]. However, the neuroscientific interpretations between different methods vary, with some methods based on phase, some on amplitude, some susceptible to volume conduction effects[9, 15, 27, 43, 44], and some sensitive to noise or signal loss[11], etc. This study utilized simulated clean scalp EEG and normative data. Therefore, the varying outcomes of the five coupling measures (COH, iCOH, PLV, PLI, and AEC) are most likely attributable to whether they are influenced by volume conduction. From **Fig. 3** and **Fig. 4**, it can be observed that in brain networks constructed by functional connectivity measures influenced by volume conduction, there exists a significant correlation between the average values of connection weights and the skewness or kurtosis of their distributions. However, such phenomenon is not evident in brain networks constructed by functional connectivity measures unaffected by volume conduction. The highly interdependent nature of scalp EEG signals is primarily attributed to volume conduction[45], which may result in MCW being able to reflect skewness and kurtosis to a certain extent. Interestingly, SE seems unaffected by volume conduction and can be well characterized by MCW (**Fig. 5**).

**Fig. 6** and **Fig. 7** display the results of constructing brain networks using COH and iCOH based on normative data, and computing the network distribution skewness, kurtosis, and SE, providing further validation of the simulation. The results support the conclusions drawn from the simulation, showing that the network exhibits a right-skewed distribution, and volumetric conduction impacts the characterization of MCW with skewness/kurtosis. From **Fig. 6** and **Fig. 7**, it can be observed that in the four frequency bands, the MCW is larger in the alpha band, consistent with prior research findings, further demonstrating that the network exhibits frequency-dependent[30, 46]. Nonetheless, the distribution of network connection weights may not be influenced by this nature.

The outlook and the limitation are as follows: (1) Considering the uncertainty of preprocessing standards and the complexity of noise situations in various EEG databases[22], using normative data for validation is adopted. Normative data only include power spectra and cross spectra information used for calculating COH and iCOH, which leads to insufficient validation for simulation. (2) The normative data cannot represent clean EEG, leading to potential errors in our results. However, the normative data underwent rigorous manual screening and quality control, which likely minimized the errors. (3) The distribution of functional connection weights at the source level has not been explored. The source-level functional connectivity may exhibit increased inhibition[12], potentially leading to a more right-skewed, leptokurtic, and uneven distribution, as inferred from the research conducted by Adamovich et al[16]. (4) This study focusing on EEG can be extended to the other electrophysiological modalities such as MEG, fNIRs.

## 5. Conclusion

In this study, we constructed brain networks using five coupling metrics and quantified the distribution patterns of network connection weights using skewness, kurtosis, and Shannon entropy, validated on normative data. Our findings indicate that resting-state network connection weights exhibit a right-skewed distribution, unaffected by channel density and coupling metrics. The distribution of resting-state network connection weights is uniform, influenced by coupling metrics, possibly attributed to whether coupling metrics are affected by volume conduction, yet the significant correlation between average connection weight and Shannon entropy remains unaffected. Networks constructed by coupling metrics unaffected by volume conduction exhibit significant correlations between average connection weight and skewness/kurtosis. There are differences in the distribution of network connection weights between frequency bands, but they are minimal. Our results serve as a foundation for better understanding and exploring potential neuroscientific information and can be applied in neuroscience projects.

## Funding

The authors did not receive support from any organization for the submitted work.

## Data availability

The power spectrum, cross-spectrum, and participant information included in the normative data can be found at https://doi.org/10.7303/syn26712979.

## Declarations

### Conflict of interest

The authors declare that they have no known competing financial interests or personal relationships that could have appeared to influence the work reported in this paper.